\documentclass{article}

\usepackage{microtype}
\usepackage{graphicx}
\usepackage{subfigure}
\usepackage{booktabs} 

\usepackage{hyperref}

\usepackage[accepted]{icml2020}

\icmltitlerunning{Machine Learning Explainability for External Stakeholders}

\begin{document}

\twocolumn[
\icmltitle{Machine Learning Explainability for External Stakeholders}



\icmlsetsymbol{equal}{*}

\begin{icmlauthorlist}
\icmlauthor{Umang Bhatt}{pai,cam}
\icmlauthor{McKane Andrus}{pai}
\icmlauthor{Adrian Weller}{cam,tur}
\icmlauthor{Alice Xiang}{pai}
\end{icmlauthorlist}

\icmlaffiliation{pai}{Partnership on AI, San Francisco, United States}
\icmlaffiliation{cam}{Department of Engineering, University of Cambridge, Cambridge, United Kingdom}
\icmlaffiliation{tur}{The Alan Turing Institute, London, United Kingdom}

\icmlcorrespondingauthor{Umang Bhatt}{usb20@cam.ac.uk}

\icmlkeywords{Machine Learning, ICML}

\vskip 0.3in
]



\printAffiliationsAndNotice{} 

\begin{abstract}
As machine learning is increasingly deployed in high-stakes contexts affecting people's livelihoods, there have been growing calls to ``open the black box'' and to make machine learning algorithms more explainable. Providing useful explanations requires careful consideration of the needs of stakeholders, including end-users, regulators, and domain experts. Despite this need, little work has been done to facilitate inter-stakeholder conversation around explainable machine learning. To help address this gap, we conducted a closed-door, day-long workshop between academics, industry experts, legal scholars, and policymakers to develop a shared language around explainability and to understand the current shortcomings of and potential solutions for deploying explainable machine learning in service of transparency goals.
We also asked participants to share case studies in deploying explainable machine learning at scale.
In this paper, we provide a short summary of various case studies of explainable machine learning, lessons from those studies, and discuss open challenges. 
\end{abstract}
\section{Overview}
\label{S:1}

In its current form, explainable machine learning (ML) is \textbf{not} being used in service of transparency for external stakeholders. Much of the ML research claiming to explain how ML models work has yet to be deployed in systems to provide explanations to end users, regulators, or other external stakeholders~\cite{bhatt2020explainable}. Instead, current techniques for explainability (hereafter used interchangeably with explainable ML) are used by internal stakeholders (i.e., model developers) to debug models ~\cite{ribeiro2016should,lundberg2017unified}. To ensure explainability reaches beyond internal stakeholders in practice, the ML community should account for how and when external stakeholders want explanations. As such, the authors of this paper worked 
with the Partnership on AI (a multi-stakeholder research organization with partners spanning major technology companies, civil society organizations, and academic institutions)
to bring together academic researchers, policymakers, and industry experts at a day-long workshop to discuss challenges and potential solutions for deploying explainable ML at scale for external stakeholders.

\subsection{Demographics and Methods}
33 participants from five countries, along with seven trained facilitators to moderate the discussion, attended this workshop. Of the 33 participants, 15 had ML development roles, 3 were designers, 6 were legal experts, and 9 were policymakers. 15 participants came from for-profit corporations, 12 came from non-profits, and 6 came from academia. 
First, participants were clustered into 5- or 6-person groups, with representation from different expertise in each group, wherein they discussed their respective disciplines' notions of explainability and attempted to align on common definitions. 
Second, participants were separated into domain-specific groups, each with a combination of domain experts and generalists, to discuss (i) use cases for, (ii) stakeholders of, (iii) challenges with, and (iv) solutions regarding explainable ML. The domains discussed were finance (e.g., employee monitoring for fraud prevention, mortgage lending), healthcare (e.g., diagnostics, mortality prediction), media (e.g., misinformation detection, targeted advertising) and social services (e.g., housing approval, government resource allocation). 

\subsection{Definitions}
``Explainability'' is ill-defined \cite{lipton2018mythos}; as such, in the first part of the workshop, the interdisciplinary groups were asked to come to a consensus definition of explainability. Below are some definitions provided by participants.
\begin{itemize}
    \item Explainability gives stakeholders a summarized sense of how a model works to verify if the model satisfies its intended purpose. 
    \item Explainability is for a particular stakeholder in a specific context with a chosen goal, and aims to get a stakeholder's mental model closer to a model’s behavior while fulfilling a stakeholder’s explanatory needs. 
    \item Explainability lets humans interact with ML models to make better decisions than either could alone.
\end{itemize}

All definitions of explainability included notions of context (the scenario in which the model is deployed), stakeholders (those affected by the model and those with a vested interest in the model's explanatory nature), interaction (the goal the model and its explanation serve), and summary (the notion that “an explanation should compress the model into digestible chunks”). Therefore, explainability loosely refers to tools that empower a stakeholder to understand and, when necessary, contest the reasoning of model outcomes. 

One policymaker noted that ``the technical community’s definition of explainable ML [is] unsettling,'' since explainable ML solely focuses on exposing model innards to stakeholders without a clear objective. Explainable ML does not consider the broader context in which the model is deployed. For a given context, the ML community's treatment of explainability fails to capture \textit{what is being explained, to whom, and for what reason}? One academic suggested that ``intelligibility could capture more than explainability;'' encapsulating explainability, interpretability, and understandability, intelligibility captures all that people can know or infer about ML models \cite{zhou2020different}. 

In the subsequent two sections, we discuss emergent themes of the domain-specific portion of the workshop. In Section~\ref{sec:com}, we discuss the need for broader community engagement in explainable ML development. In Section~\ref{sec:dep}, we outline elements of deploying explainable ML at scale.

\section{Designing Explainability}
\label{sec:com}
The first salient theme noted by participants was the lack of community engagement in the explainable ML development process. Community engagement entails understanding the context of explainable ML deployment, evaluation of explainable ML techniques, involvement of affected groups in development, and education of various stakeholders regarding explainability use and misuse.

\subsection{Context of Explanations}
\label{sec:context}
Given the context of the deployed model, an explanation helps stakeholders interpret model outcomes based on additional information provided (e.g., understanding how the model behaves, validating the predictability of the model's output, or confirming if the model's ``reasoning aligns with the stakeholder's mental model'')~\cite{ruben2015explaining}.

Each stakeholder may require a different type of transparency into the model. Expanding the ML community's understanding of the needs of specific stakeholder types will allow for model explanations to be personalized. The notion of a good explanation varies by stakeholder and their relevant needs \cite{arya2019one,miller2019explanation}.

To further probe these contexts and understand what stakeholders actually need from explanations, many participants pointed to the need for explainable ML to incorporate expertise from other disciplines. Introducing researchers from human-computer interaction and user experience research as well as bringing in community experts were seen as ways to enable participatory development and to ensure the applicability of explainable ML methods.

Another dimension of context that participants noted is that ML systems represent a chain of models, data, and human decisions \cite{lawrence2019data}, or, in other words, a distinctly sociotechnical system (See \cite{selbst2019fairness} for a summary of common issues faced with sociotechnical systems). An organization that has many models in production will require different levels and styles of transparency for each stakeholder to operate cohesively. 
At times, these transparency requirements can be just a matter of disclosure of the process. Though, making that information available could be nontrivial \cite{raji2019ml,arnold2019factsheets,gebru2018datasheets,mitchell2019model}.

\noindent \textbf{Takeaway}: Explainability tools cannot be developed without regard to the context in which they will be deployed.

\subsection{Evaluation of Explanations}
As part of deploying technical explainability techniques in different contexts, practitioners described a need for clarity on how to evaluate explainable ML's effectiveness. Given the wide range of potential uses for explainability, it is not clear how stakeholders should agree upon or test for the desirable properties of an explanation \cite{doshi2017towards}. Quantitative evaluation of explanations, like in \cite{hase2020evaluating,bhatt2020evaluating}, are a starting point for this work, and qualitative studies of how to combine models and explanations with stakeholders in a decision making process \cite{bansal2019beyond} are a critical next step. Even amongst researchers focused on explainable ML, there is no consensus on how to evaluate an explanation, let alone an understanding of which explanation techniques are good at helping stakeholders achieve their goals in specific contexts.

Participants discussing the role of explainable ML in journalism and social media pointed to the difficulty of understanding how users understand and internalize explanations they are given about mis-/dis-information. Cognitive biases such as the back-fire effect \cite{peter2016debunking}, where users double down on prior beliefs when confronted with contradictory evidence, can completely invert the intended effect of explaining why an article is deemed inaccurate. Attempts at explanation evaluation, especially automated, quantitative evaluations, can very easily miss these more contextual elements \cite{doshi2017towards}.

To effectively evaluate explanations, participants wanted rigorous human evaluation of explainability; to date there are few examples of this \cite{poursabzi2018manipulating}. Participants called for more interdisciplinary collaboration by bringing in experts from human-computer interaction, user experience research, and socially-oriented disciplines to help establish explanation evaluation in specific contexts.


\noindent \textbf{Takeaway}: When developing explainable ML, clarity in how organizations evaluate explainability algorithms or how individuals measure explanation utility is essential.

\subsection{Appropriate Design for Affected Groups}
\label{sec:involving}
As discussed in Section~\ref{sec:context}, a key component of explainability is answering the question of \textit{what} is being explained \textit{to whom}. When designing an explainable ML system, those deploying the system have a decision to make on how thoroughly they attempt to understand the breadth of relevant stakeholder needs. Below we outline a few salient areas where better understanding affected groups could go a long way in improving explainable ML.

Participants pointed to scenarios where communities might have disparate capacities to engage with explainable ML. One scenario posed was the case of an apartment rental application tool, which ought to explain to applicants why they may be denied. Participants thought it was likely that brokers and applicants with institutional knowledge would be able to modify future applications to improve their chance of success, whereas already disenfranchised applicants would be stuck in cycles of rejection. One participant proposed using simpler models that produce actionable explanations as a way to reduce this effect. Understanding these differential responses in non-theoretical cases, however, will likely require designing and evaluating systems directly alongside impacted communities. 

A different aspect of explainability entails ``being specific enough that you are giving actual meaningful information about how [input] data is being used,'' as one participant noted. In the healthcare domain, protections in this vein have already been codified into law. HIPAA \cite{hipaa} in the US and GDPR \cite{gdpr} in Europe require confidential and transparent management of medical data. As a result of these patient protections, participants noted that any type of explanation using the training data is unlikely to be deployed in this domain. For other application areas, however, it is less likely that such stringent data protections will apply, leaving it to organizations to decide how protected and transparent individual data use should be.

A follow on to the previous issue is determining what data should be used at all. In many settings, ML models are trained on potentially irrelevant data or sensitive data that might raise privacy concerns. One potential benefit of explainable ML is that issues of data misuse can be more directly addressed. Explanation recipients, whether they are credentialed experts (e.g., doctors) or the actual subjects of decisions (e.g. rental applicants), likely have a prior understanding of which attributes should be relevant to the decision being made. By having explanations explicitly mention the attributes being used in decision making, these stakeholders can be empowered to contest privacy encroachments and to challenge questionable decisions. 

\noindent \textbf{Takeaway}: Including stakeholders in the development of explanations and striving to better understand stakeholder needs can prevent preferential treatment and data misuse.

\subsection{Stakeholder Education}
Understanding how to educate stakeholders regarding explainability is key to its widespread adoption. One participant noted that ``data scientists are aware of explainable ML but are clueless about how to use'' it: data scientists have not been provided with a best practices framework for choosing which explanation technique to deploy in various contexts or how to do so successfully \cite{kaur2019interpreting}.
One participant from a financial institution stated that ``data scientists are not demanding education on how to use [explainable ML], since they optimize their career and will only focus on [explainable ML] when they have to or when they know that it will be brought up to them.'' In certain domains, explainability requirements are \textit{top-down} (regulators are mandating a specific form of explanation from models); however, widespread adoption of explainable ML will likely require \textit{grassroots} education of data scientists, who are aware of the context in which the model is deployed.

One issue in explainable ML stakeholder education is ensuring stakeholders are aware of the limitations of post-hoc explanations. A post-hoc explanation provides insight into a pre-trained model in the form of important features, important training points, or decision boundary analysis. One participant noted that ``feature importance methods might be able to provide [transparency], but if they are post-hoc explanation methods, we do not know if we can trust that the explanation reflects reality. Post-hoc [explanations] are limiting and are loosely termed an explanation; they might not be a useful justification of the reality of what is going on in the model internally.'' 
\cite{weller2019transparency} notes that transparency of ML models can allow malicious attackers to provide deceptive information as an ``explanation;'' recent work has concluded that feature importance techniques can be manipulated to fool end users \cite{slack2020fooling} and to conceal model unfairness \cite{dimanov2020you}. Informing stakeholders of explainable ML's potential to mislead unintentionally or to deceive purposefully is critical.

Another participant from the healthcare domain noted that clinicians have background knowledge and  training in making diagnoses, but for the clinician to feel comfortable vetoing a diagnostic model, the clinician must be aware of the model's failure modes and understand how the model works. Sometimes there can be no time for clinicians to get the training required to do this translation (or no space in the medical school curriculum).
There may be an emerging career where one has specialties in clinical training and in ML, almost like analytic translators who are able to translate model behavior to clinicians and who understand the nuances of the model’s specification (similar to radiologists today). 
Future research could address how to integrate explainability into ML curricula and into curricula of the stakeholders making decisions based on model outputs.

In addition to data scientist and domain expert education, public education around ML and explainability is crucial. People deserve to know when an ML model is being used in a decision regarding them. ``The techies need to hear what people are afraid of... most people do not know they are interacting with AI'' stated one policymaker.  Public education would require a common vocabulary that is simple for non-experts to understand and avoids obscure jargon.

\noindent \textbf{Takeaway}: Developing curricula for stakeholders will encourage thoughtful adoption of explainable ML, while accounting for differences in expertise and bandwidth.


\section{Deploying Explainability}
\label{sec:dep}
In addition to engaging with the community around developing explainability tools, participants also discussed the many nuances of deploying such tools in practice. 

\subsection{Uncertainty alongside Explanations}
Existing literature limits their view of post-hoc explanations to feature importance \cite{ribeiro2016should,lundberg2017unified,davis2020network}, sample importance \cite{koh2017understanding,yeh2018representer,khanna2019interpreting}, or counterfactual reasoning \cite{wachter2017counterfactual,dhurandhar2018explanations,ustun2019actionable}; however, it is also important to consider the uncertainty associated with model predictions. Some participants noted that predictive uncertainty can be complementary to an explanation.


One participant from a healthcare organization noted that some diseases are more well-understood than others. When deploying diagnostic decision support tools for predicting which disease a patient has, clinicians need to understand how confident the model is for the suggested prediction. Ideally, the clinicians should decide the threshold at which the model can safely make a prediction of a rare disease. Uncertainty within the model ought to be higher for rare diseases than for common ones, but in practice it is difficult to quantify and communicate predictive uncertainty. Rigorously measuring and exposing uncertainty alongside an explanation could be useful to clinicians who can leverage their expertise to make informed decisions.

As this participant's experience indicates, predictive uncertainty is difficult to accurately measure in practice. Many classification models in use today provide ``class probabilities,'' which represent how likely each class is relative to other classes. Usually, the highest class probability for a datapoint is taken to be its classification. As such, the maximum class probability is often referred to as the model's confidence. However, maximum class probability has been shown to be poorly correlated with the true class probability in deep learning models \cite{guo2017calibration}. Class probabilities for datapoints the model has not seen before (usually called out-of-distribution data) are unreliable \cite{snoek2019can}. When predictive uncertainty accompanies an explanation, class probabilities must be calibrated with empirical outcomes: average confidence should not exceed average accuracy. Numerous methods for better calibrated predictive uncertainty have been proposed \cite{kuleshov2018accurate,kumar2018trainable,corbiere2019addressing}, but it is unclear how they might interact with other strategies for improving explainability. If these methods can be reconciled, studying how to visualize and convey model confidence could make explainability more useful for external stakeholders.

Luckily, once uncertainty is accurately measured, there is a plethora of work on conveying model confidence (more generally, on communicating statistics)  \cite{spiegelhalter2017risk,hullman2019authors}. In specific situations, it may be sensible to expose this uncertainty to a human decision-maker \cite{zhang2020effect}; for example, showing a mortgage approver for which applicants (or better yet, for which of the applicant's features) a model is uncertain could help the approver know when to intervene in a automated decision-making process \cite{antoran2020getting}. 
Future research should explore the role of uncertainty in explainable ML and develop frameworks for how to expose this information to stakeholders.

\noindent \textbf{Takeaway}: Treating confidence as complementary to explanation requires the ML community to develop context-specific techniques for quantifying and communicating uncertainty to stakeholders.

\subsection{Interacting with Explanations}
Most existing post-hoc explanation techniques convey information about the model to stakeholders; however, few techniques have been developed to update a model based on the stakeholder's view of the explanation \cite{lee2020explanation,bansal2019updates} or to provide stakeholders with the ability to toggle the information in an explanation.

Explanations from ML models effectively provide evidence, and stakeholders then examine that evidence, noting if it aligns with their intuition \cite{bansal2019beyond,buccinca2020proxy}.
Stakeholders should be able to interact with the explanation to control how much information is conveyed: if a stakeholder wants less information, the explanation technique used should convey a more summarized explanation without requiring changes to the underlying model. Over time, as a model behaves as expected (in line with the stakeholder's mental model), explanations could serve as evidence for potential model failures. Perhaps, an explanation may not be required when the model displays trustworthiness via consistently accurate predictions. Stakeholders may let failures slide as model predictability builds over time, but explanation techniques must adapt to the specificity that the stakeholder desires. 

One participant from a civil society organization noted that interactive explanations, which allow stakeholders to peek inside a model's behavior, are important when deploying ML models for resource allocation by governments and when providing natural language explanations alongside predictions. However, if a front-line practitioner (i.e., a government official checking for farmer compliance) cannot override the model's prediction for a particular individual, then practitioners grow skeptical of the model's utility. Front line practitioners want to ask  questions to the model about its learned reasoning and want to provide feedback to the model in real-time. Flexible, interactive models that allow practitioners to alter trained models online to reflect practitioners' mental models are crucial.

Another participant noted that, in their organization, language models conflated Paris Hilton with Hilton Hotels and City of Paris; their organization lacked procedures for a data scientist to expose and alter these correlations to reflect reality. Future research ought to develop actionable tools for correcting a model suffering from spurious correlations and other errors exposed by a model's explanation. How to mathematically formalize the feedback received from the stakeholder regarding the explanation and how to update the model prior, in some sense, based on the feedback are open questions. Tools that enable interactions with models, documentation that enumerates implicit assumptions in model training, and interfaces that allow stakeholders to interrogate models are essential for adopting explainable ML \cite{powerPeople2014}. Model interactivity may require interpretability by design, wherein the model itself is explainable, due to the chosen model class, instead of deriving post-hoc, approximate explanations \cite{rudin2019stop}.

Some elements of explainability are indirect. Another participant noted that a clinician might want an explanation from a diagnostic model. The model itself or a post-hoc explanation technique can create an explanation for the patient that the clinician can deliver verbally in a conversation, wherein the statistical rigor (false positive rate, feature importance, etc.) is provided only if the patient asks for specificity. Explainable ML in hybrid human-machine decision-making may only be necessary up until a certain point: stakeholders need interactivity to ensure the model aligns with their own mental model. Thereafter, model predictability (reliability in prediction) matters more than model transparency.

Interaction is the keystone of shared human-machine decision making. Interacting with a model (either based on its predictions/behavior or based on its reasoning/explanation) is a way to facilitate a synergistic dialogue between humans and machines \cite{amershi2019guidelines}. In mortality prediction, clinicians need to ascertain if the model captures the goals of potential treatment plans, the preferences of patient lifestyle, and circumstances of patient history. Interactivity extends explainability beyond a one-way information transfer, such that users can exercise contestability.

\noindent \textbf{Takeaway}: Creating flexible explanation techniques that stakeholders can toggle and building models that can update based on stakeholder feedback provided will encourage adoption of explainable ML at scale.

\subsection{Behavior Changes from Explanations}
In many domains, participants noted that a key component of explanations is how actionable they are for different stakeholders. Whether this was in the case of hospitals improving their health outcomes or journalists removing references to mis-/dis-information, there are specific actions motivated by the explanation a stakeholder receives. 

As such, issues can arise when explanations do not account for how stakeholders might respond to them. In the worst case for system designers, explanations will hone in on easily modifiable characteristics (i.e. the number of friends one has on a social media account) or difficult to alter and seemingly unimportant characteristics (e.g. zip code). Participants discussed how stakeholders are likely to lose trust in the model or in the decision-making process as a whole. 

One discussed example of this, though not explainability-specific, was the ``pain-management'' component of the Medicare hospital satisfaction score. By scoring hospitals according to how well patients believed their pain was managed, the Medicare score established a perverse incentive to overly prescribe opioids and antibiotics. If suggested courses of action to improve the score were given alongside the score itself, the outcome could have been different.

Inherent to these issues is what one participant described as ``a philosophical question about the meaning and positioning of explanations.'' If by altering their pertinent attributes stakeholders are perceived as using explanations to ``game'' metrics, there is some question of how relevant those attributes really are. Going back to the explainable rental application system example from Section~\ref{sec:involving}, a key motivation for designing such a system should be informing applicants on how to become better applicants in the future. For less expert stakeholders, counterfactual explanations along significant axes were deemed the most actionable. Adopting this treatment would increase the ability of honest affected stakeholders to be correctly classified as qualified tenants or to make the sufficient changes for positive classification, reducing what has elsewhere been referred to as the ``social cost'' of the model \cite{milli2019social}. 



\noindent \textbf{Takeaway}: When designing an explainable ML tools, include how the explanations might be acted upon as a central design question. If the explanations motivate the average user to game or distrust the system, perhaps it points to the model making predictions on unfair/unimportant attributes. 

\subsection{Explainability over Time}
As model functions are made more transparent and stakeholder behavior adapts, it is likely that model performance will similarly start to shift. As stated under Goodhart's Law (as re-phrased by Marilyn Strathern, ``When a measure becomes a target, it ceases to be a good measure,'' \cite{strathern1997improving}. The key insight of this ``law'' is that once a metric is used for informing decisions, people have incentives to optimize that metric to achieve the decision they want.

Gaming is often blamed as a main source of this distributional shift, but, as mentioned in the previous section, a key component of explainability is how actionable the explanations are. As such, explainability tools should be designed with the explicit expectation that the underlying distributions are going to change. The predictability of this distributional shift can be seen as more blessing than curse, as it encourages more flexible system design and can inoculate the organization against common model failures.

One example brought up by participants was the case of a health-outcome prediction model. A theoretical patient is predicted as high risk and the doctor is given an explanation that attributes much of this risk to the patient’s weight. Given this information, the patient might lose weight to improve their prognosis. It could be that a significant risk factor correlated with weight, such as hypertension, is not reduced concomitantly with weight. If the model does not explicitly include hypertension, it is likely to underestimate the risk of a patient who has lost weight but still has hypertension. When only a few individuals make this change, overall accuracy might not drop by much. However, if encouraging weight loss becomes a standardized treatment plan for a doctor, we can expect the model’s accuracy and utility to drop if the model is not updated to reflect the new patient archetype. 

Beyond distributional shift, participants also discussed how professionals working closely with a model might adapt to it over time. Drawing from the healthcare conversation once more, one participant pointed to the trust dynamics between nurses or doctors and explainable ML. At first there is likely to be a lack of trust, but trust can grow if the tool proves accurate and useful in whatever task it was designed for. Reaching a more trusting, comfortable state, however, often means not just blindly following the tool’s recommendations, but incorporating them into the daily judgments one makes. As another participant mentioned, this means that updates to the system, even if they technically improve accuracy or the explanation quality, can cause a mismatch between model behavior and user expectations that worsens overall performance. There has been some work on the dynamics of updating ML systems in human-machine teams more generally \cite{bansal2019updates}, but explainable ML models are likely to be a unique case given the different types of interaction they allow.

\noindent \textbf{Takeaway}: Explainability tools enable adaptation by affected parties and system users, so successful deployment will require frequent accounting for these adaptations.



\section{Conclusion}
This paper outlined the findings of an interdisciplinary convening of stakeholders of explainable ML.
We found that future research around explainability could benefit from community engagement in explainable ML development and from thoughtful deployment of explainable ML.
Understanding the context in which an explanation is used, evaluating the explanation accordingly, involving affected stakeholders in development, and educating stakeholders on explainability are keys to the adoption of explainable ML. 
While deploying explainable ML, stakeholders should consider whether the uncertainty of the underlying model affects explanations, how stakeholders will interact with explanations, how stakeholders behavior will change due to an explanation, and whether stakeholders require transparency in the form of explanations after repeated interactions with models.
If future research involves the community in development and cautiously deploys explainable ML, explainability can be used in the service of transparency goals.
We urge researchers to engage in interdisciplinary conversations with external stakeholders. Input from external stakeholders will increase the utility of explainable ML beyond the ML community.

\bibliography{paper}
\bibliographystyle{icml2020}

\end{document}